\documentclass[a4paper,11pt]{article}
\usepackage{times}
\usepackage{datetime}
\usepackage{pos}
\usepackage{multicol}
\usepackage{float}

\title{Expected Performance of the EUSO-SPB2 Fluorescence Telescope}
 \ShortTitle{Expected Performance EUSO-SPB2 FT}

\author*[a]{G. Filippatos}
\author[b]{M. Battisti}
\author[b,c]{M. Bertaina}
\author[c]{F. Bisconti}
\author[d]{J. Eser}
\author[e]{C. Heaton}
\author[f]{G. Osteria}
\author[a]{F. Sarazin}
\author[a]{L. Wiencke}

\affiliation[a]{Colorado School of Mines, Golden, USA}
\affiliation[b]{Universita’ di Torino, Torino, Italy}
\affiliation[c]{Istituto Nazionale di Fisica Nucleare, Turin, Italy}
\affiliation[d]{University of Chicago, Chicago, USA}
\affiliation[e]{The Pennsylvania State University, State College, USA}
\affiliation[f]{Istituto Nazionale di Fisica Nucleare, Naples, Italy}

\forColl{JEM-EUSO}

\emailAdd{gfilippatos@mines.edu}

\abstract{The Extreme Universe Space Observatory Super Pressure Balloon 2 (EUSO-SPB2) is under development, and will prototype instrumentation for future satellite-based missions, including the Probe of Extreme Multi-Messenger Astrophysics (POEMMA).
EUSO-SPB2 will consist of two telescopes. The first is a Cherenkov telescope (CT) being developed to identify and estimate the background sources for future below-the-limb very high energy (E>10 PeV) astrophysical neutrino observations, as well as above-the-limb cosmic ray induced signals (E>1 PeV).
The second is a fluorescence telescope (FT) being developed for detection of Ultra High Energy Cosmic Rays (UHECRs).
In preparation for the expected launch in 2023, extensive simulations tuned by preliminary laboratory measurements have been performed to understand the FT capabilities.
The energy threshold has been estimated at 10$^{18.2}$ eV, and results in a maximum detection rate at 10$^{18.6}$ eV when taking into account the shape of the UHECR spectrum.
In addition, onboard software has been developed based on the simulations as well as experience with previous EUSO missions.
This includes a level 1 trigger to be run on the computationally limited flight hardware, as well as a deep learning based prioritization algorithm in order to accommodate the balloon’s telemetry budget.
These techniques could also be used later for future, space-based missions.}

\FullConference{37$^{\rm{th}}$ International Cosmic Ray Conference (ICRC 2021)\\
		July 12th -- 23rd, 2021\\
		Online -- Berlin, Germany}

\begin{document}
\maketitle

\section{Introduction}\label{secs:Background}

The Extreme Universe Space Observatory Super Pressure Balloon 2 (EUSO-SPB2) is a step along the path towards a space-based Ultra High Energy Cosmic Rays (UHECRs) observatory.
Building on the experience of previous EUSO balloon missions, EUSO-Ballon \cite{EUSO-Ballon} and EUSO-SPB1 \cite{SPB1}, EUSO-SPB2 aims to serve as a proof of concept by being the first instrument to measure extensive air showers (EAS) from above via fluorescence observations.
In addition, the EUSO-SPB2 fluorescence telescope (FT) will prototype instrumentation that is planned to be flown on future space-based missions such as POEMMA \cite{POEMMA} or K-EUSO \cite{K-EUSO}.

\begin{figure}[H]
\centering
\begin{minipage}{.3\textwidth}
  \centering
  \includegraphics[width=\textwidth]{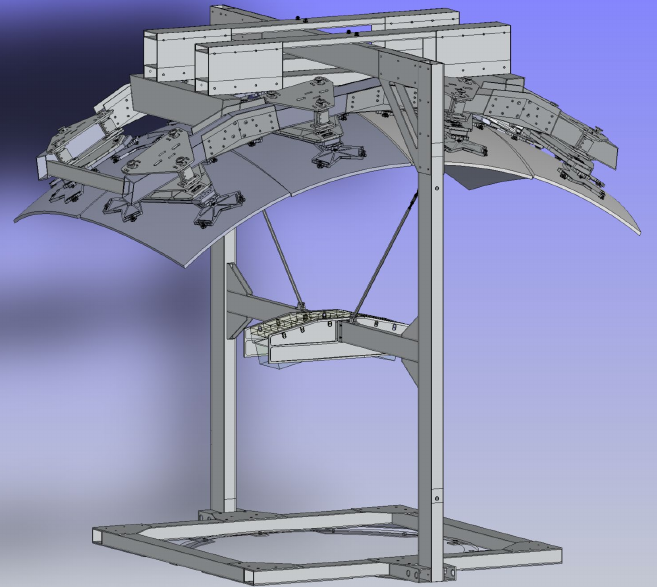}
  \caption{CAD rendering of the FT.}
  \label{fig:Telescope}
\end{minipage}%
\hspace{0.5cm}
\begin{minipage}{.3\textwidth}
  \centering
  \includegraphics[width=\textwidth]{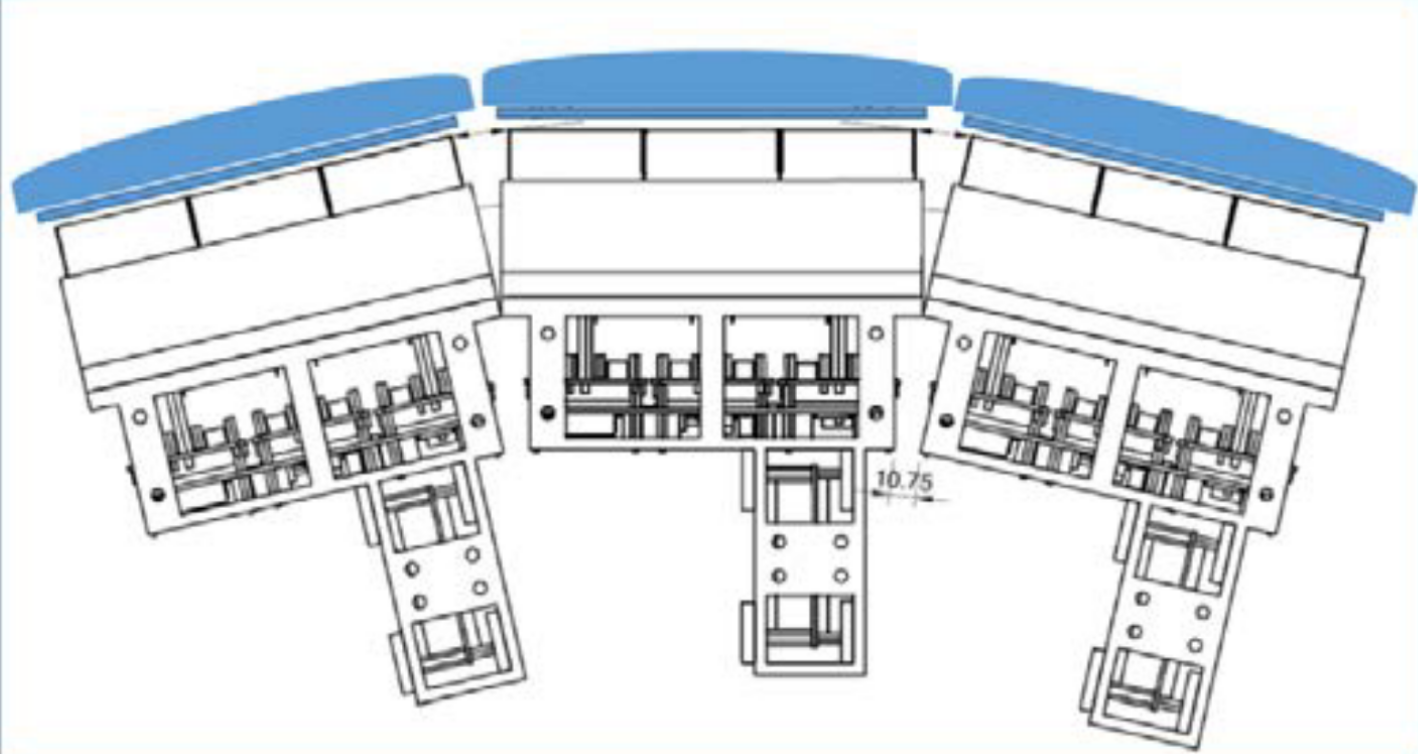}
  \caption{Schematic image of the three PDMs as they will be arranged.}
  \label{fig:PDMs}
\end{minipage}%
\hspace{0.5cm}
\begin{minipage}{.3\textwidth}
  \centering
  \includegraphics[width=\textwidth]{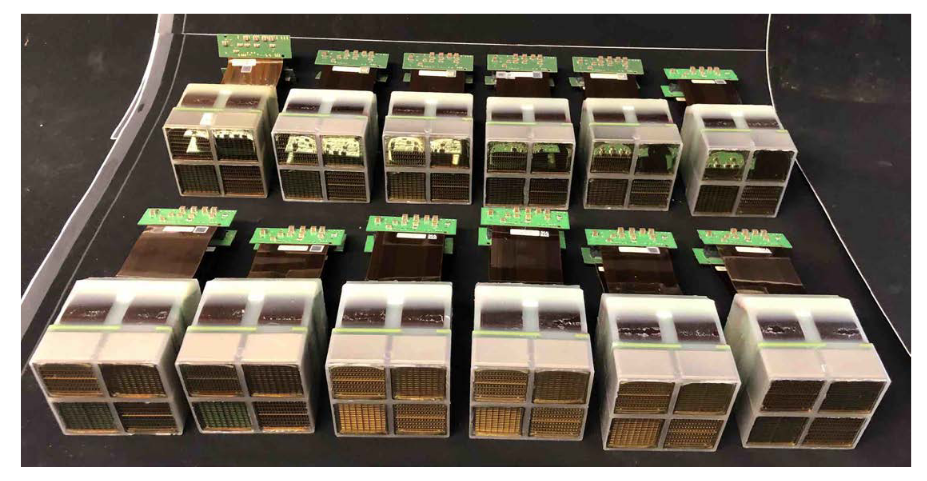}
  \caption{Photograph of 12/27 ECs after assembly.}
  \label{fig:ECs}
\end{minipage}
\end{figure}

The FT will consist of three photo-detection modules (PDMs), of which EUSO-Balloon and EUSO-SPB1 consited of one.
These PDMs contain nine elementary cells (ECs) which each include four 8x8 multi-anode photo-multiplier tubes (MAPMTs), for a total 2,304 pixels per PDM, each capable of photo-electron counting.
A CAD rendering of the telescope is shown in Figure \ref{fig:Telescope}, a schematic image of the PDMs is shown in Figure \ref{fig:PDMs} and the actual ECs are shown in Figure \ref{fig:ECs}.
Each pixel is 3mm x 3mm, and has an integration time of 1 $\mu$s, with a double pulse resolution of 6 ns.
Light is focused onto the focal surface by a Schmidt telescope consisting of six mirror segments, with a field of view of ~36$^\text{o}$x12$^\text{o}$ as well as an aspheric corrector plate to account for spherical aberrations.

EUSO-SPB2 will be flown on a NASA super pressure balloon \cite{SPB} at an altitude of 33 km.
Super pressure balloon flights can be long duration, lasting up to 100 days.
Launching from Wanaka NZ, at -45$^\text{o}$ latitude, these balloon flights utilize a fast stratospheric air circulation that develops twice a year at about 33 km (7 mbar) above the southern ocean.
This airflow pattern allows the balloon to travel around the globe at a constant latitude with a somewhat predictable flight path.
The balloon will spend most of its time over the ocean, and enough time over land to allow for a possible recovery of the payload at the end of the flight.
Recovery is not guaranteed, the only data that is guaranteed is what can be downloaded during flight.
Therefore, a robust data prioritization scheme needs to be developed to ensure that the most important data is transmitted to ground.

There is limited telemetry available during flight, which will be shared among two telescope, ancillary devices and  housekeeping data.
Only about 1\% of data recorded by the FT will be downloaded during flight.
Even so, a very small fraction of recorded data will contain signal from EAS, roughly 3x10$^{-3}\%$.
Therefore it is necessary to classify events using the on-board software.
Further, this classification will need to be performed subject to the computational constraints of the payload.
This problem will be similar in satellite-based missions as well.
We have elected to use a convolutional neural network approach trained on simulated data, and pre-flight field test data when available, to classify events.

Prior to this selection of events for download, the instrument is read out based on a hardware level 1 trigger.
This process needs to be done in real time in order to decide whether or not to readout data.
Therefore, a computationally simple method is required.
Since EUSO-SPB2 will be flown at a much lower altitude than satellite-based missions, a new level 1 trigger has been developed to better fit the different observational scenario.
The details of this trigger are described in Section \ref{secs:Trigger}, the impact this has on the expected event rate is described in Section \ref{secs:EventRate} and the details of the prioritization algorithm are described in Section \ref{secs:CNN}.

\section{Level 1 Trigger Development}\label{secs:Trigger}

The first step in the FT data acquisition (DAQ) process is recognizing a possible event of interest.
This is done with a level 1 trigger.
In order to identify a potential EAS signature, rather than using a simple threshold trigger where only an excess of signal over background is observed, a more complex trigger has been developed.
The general idea is to look for multiple clusters of excess signal within a certain time window, in a manner that is general enough to capture all possible geometries of EAS passing in front of the telescope, while being specific enough to reject noise resulting from Poisson fluctuations of the light background.

Each PDM triggers independently, the first step in the trigger algorithm is to segment a PDM from a 48x48 pixel array into a 24x24 macro-pixel array by grouping adjacent clusters of 4 pixels with no overlap.
The motivation for this is that using pixels designed for a satellite borne instrument at 525 km on an instrument being flown at 33 km leads to the majority of EAS, depending on the inclination of the shower, crossing more than one pixel per integration even with perfect optics.
By analyzing macro-pixels, the relative excess of signal per macro-pixel is larger than if the pixels were analyzed individually.
Macro-pixels are then analyzed to be above threshold if the sum of the signal in all four pixels for a given integration is above $S_{\text{Thresh}}$ which is calculated as shown in Equation \ref{eqs:TriggerThreshold}
\begin{equation}\label{eqs:TriggerThreshold}
S_{\text{Thresh},i} = n_{\sigma} \sqrt\lambda_i + \lambda_i
\end{equation}
where $\lambda_i$ is the average count rate for macro-pixel $i$, averaged over 16,384 $\mu$s and calculated every 0.5 seconds.
If the signal in a macro-pixel satisfies $S_i > S_{\text{Thresh},i}$ it is considered "hot".
Next, clusters of "hot" macro-pixels are searched for within a 3x3x3 (x,y,t) grid.
Each of these grids can be defined by the central macro-pixel, with macro-pixels on the edge of the PDM being excluded.
If the number of macro-pixels that are "hot" in a given grid is $>n_{\text{hot}}$ then that grid is considered "active".
When a given number $n_{\text{active}}$ grids are active within a defined time frame $l$ a trigger is issued.
Once a trigger is issued, the 64 frames before and 64 frames after are read out on all three PDMs.

As can be seen, the trigger is controlled by the parameters $n_{\sigma},n_{\text{hot}},n_{\text{active}},l$ which can be tuned during flight to control the trigger rate and avoid saturating the DAQ electronics.
A wide phase space of these parameters was investigated to find an optimal configuration that would yield a trigger rate manageable by our instrument.
The configuration yielding the highest event rate with a tolereable trigger rate of ~1 Hz was found to be: $n_{\sigma}=5.0,n_{\text{hot}}=2,n_{\text{active}}=34,l=20 \text{ frames}$.

Overall, this trigger has several distinct advantages over previous EUSO trigger implementations \cite{JemEUSO-Trigger}.
The primary advantage is that thresholds are set locally on a per macro-pixel basis rather than a per MAPMT basis.
A result of this is a single hot pixel, will not blind the entire MAPMT.
Additionally, clustering of signal is searched for in the entire PDM rather than in an individual MAPMT.
This results in EAS that have a lower maximum brightness being detected by virtue of their signal persisting across a large section of the camera.
Lastly, by clustering pixels into macro-pixels excess signal is searched for in a manner that more closely resembles the physical signal of an EAS in the camera.

\section{Expected Event Rate}\label{secs:EventRate}

\begin{figure}[H]
\centering
  \centering
  \includegraphics[width=\textwidth]{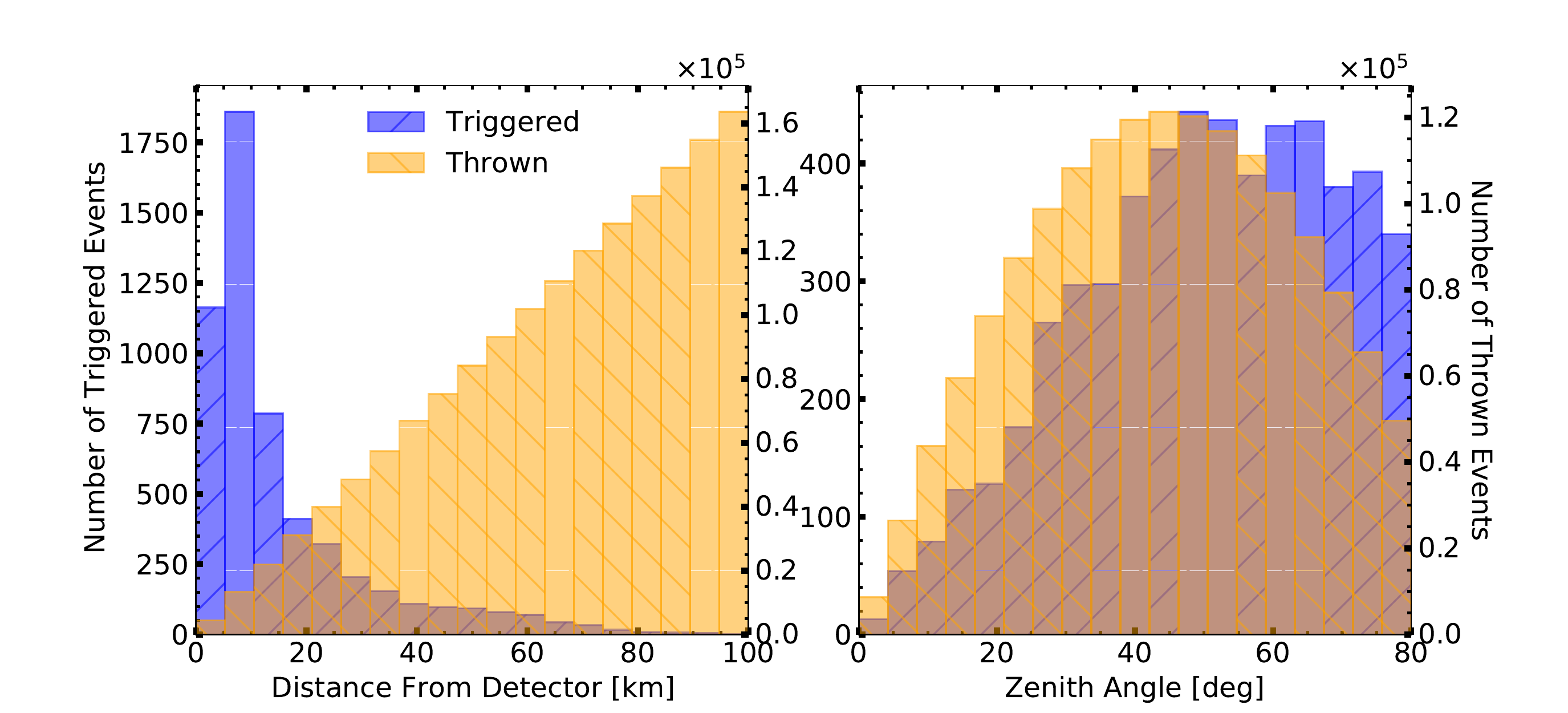}
  \caption{Distribution of core locations, shown as distance from the point directly below the detector to core location at sea level, of triggered showers and thrown showers (left).
  Distribution of zenith angles of triggered showers and thrown showers (right). }
  \label{fig:DistZenith}
\end{figure}

In order to estimate the observation event rate of the instrument, a large scale Monte-Carlo approach is used.
The possible geometries are less trivial than for ground based detectors, as a shower with a core location (the point where the shower axis crosses sea level) far away from the detector may still pass through the field of view while depositing energy.
For this reason, an over-sized area needs to be sampled.
Showers are thrown to have a core location uniformly distributed over a 31,415 km$^2$ disk on ground, centered around the detector, the projected FoV of which only covers roughly 36 km$^2$.
These showers sample a zenith angle distribution that is flat in $\sin\theta\cos\theta$, in order to account for the projected area of the disk which is perpendicular to the shower axis, and an azimuth distribution that is uniform in $\phi$.
The distribution of core locations and zenith angles for both thrown and triggered showers is shown in Figure \ref{fig:DistZenith}.
For this study, the sampled energy bins are evenly spaced in $\log_{10}(\text{E}/\text{eV})$, with 20 bins ranging from $10^{17.8}$ eV to $10^{19.7}$ eV with 80,000 showers thrown per energy bin.
The background is based on the most realistic estimate from previous experiments and accounts for the non-uniformity of the MAPMTs, simulated signals from direct cosmic ray hits and the expected average airglow as a function of observational angle of the pixels \cite{NightGlow}.
Using the energy spectrum as measured by the Pierre Auger Observatory \cite{AugerSpectrum}, the number of triggered events is converted into an expected event rate as shown in Equation \ref{eqs:EventRate}.
\begin{equation}\label{eqs:EventRate}
R(E_{i})= \bigg(\frac{\text{N}_\text{Trigger}}{\text{N}_{\text{Thrown}}} \bigg) A \Omega \int_{E_i-\Delta E/2}^{E_i+\Delta E/2} J(E)dE .
\end{equation}
Where $E_i$ is the energy bin, $R(E_{i})$ is the number of observed events per hour, $A$ is the area sampled $\Omega$ is the solid angle sampled $\Delta E$ is the width of the energy bin and $J(E)$ is the energy spectrum.

\begin{figure}
  \centering
  \includegraphics[width=0.9\textwidth]{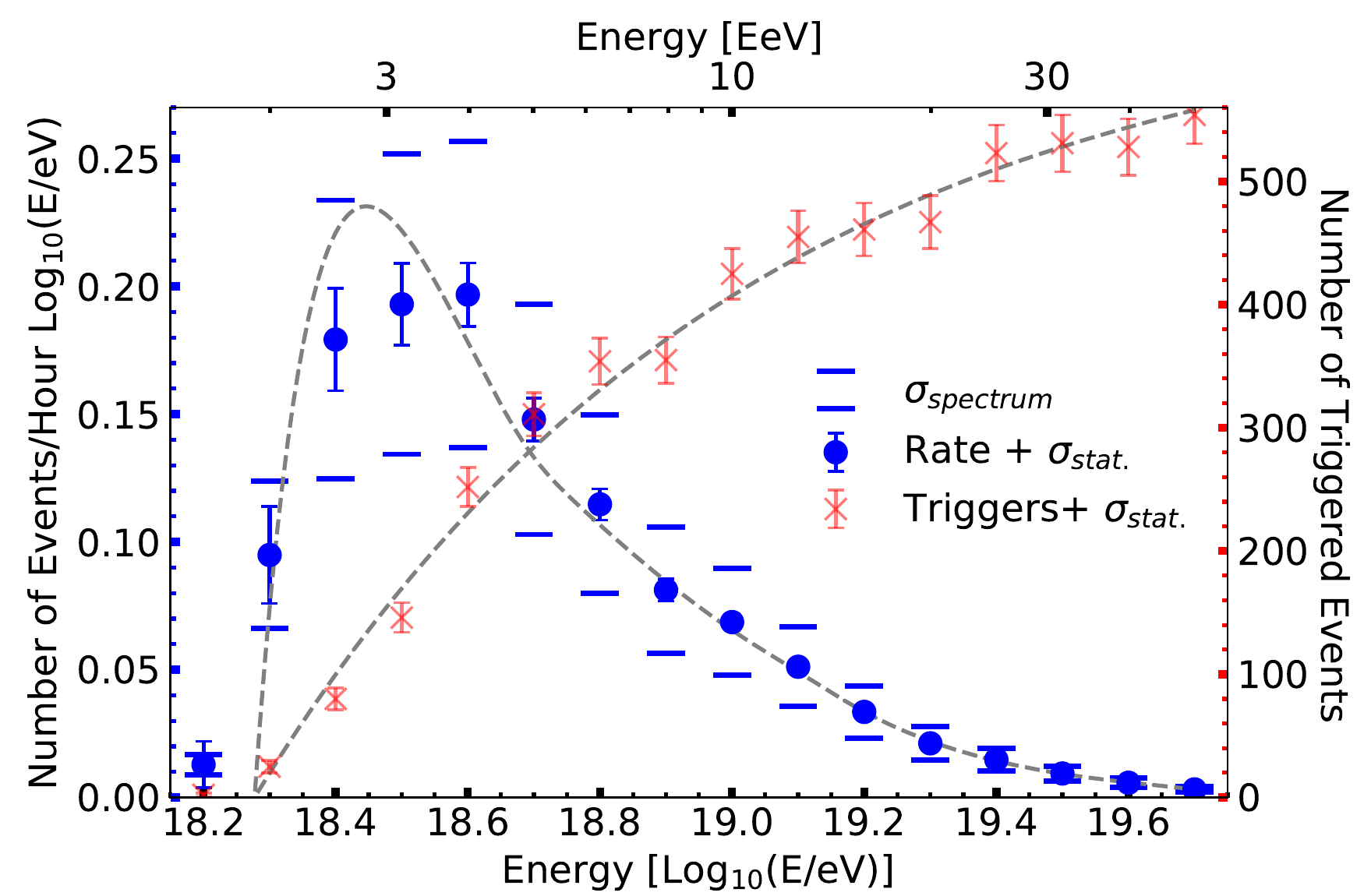}
  \caption{The number of triggered showers per energy bin (red).
  Fit to an activation function of the form $a(1-\exp{(-(x+b)/c)})$.
  The estimated event rate as a function of shower energy (blue).
  Normalized to be differential with respect to $\log_{10}(\text{E}/\text{eV})$.
  Statistical error bars $\sigma_{\text{stat.}}$ shown are Poissonian errors on the number of triggered events.
  Uncertainty of the spectrum as measured by Auger shown as $\sigma_{\text{spectrum}}$.}
  \label{fig:EventRate}
\end{figure}

The resulting event rate as a function of shower energy is shown in Figure \ref{fig:EventRate}.
The curve is normalized to be a differential energy spectrum.
An activation function is used to fit the raw number of triggered events and then converted to an event rate following Equation \ref{eqs:EventRate}.
Summing the contributions from all energy bins yields $0.12\pm0.01\pm0.04$ events per hour, hence about 0.6 events per night assuming a 5 hour observation window.
The systematic uncertainty in the energy spectrum measured by Auger is $\approx 30$ to 40\%, providing the largest source of uncertainty in the expected event rate.
As a result the goal of observing the first sub-orbital EAS via fluorescence is well within reach considering a 14 day flight, the minimum duration expected.

\section{Prioritization Algorithm}\label{secs:CNN}

Due to the limited telemetry bandwidth on a super pressure balloon flight, and the wide range of data that need to be sent down, we estimate that roughly 1\% of the recorded events can be downloaded during flight.
Triggering at about 1 Hz yields around 15,000 events per night, of which a very small fraction will be EASs.
In order to fit within the constraints of our telemetry budget, a method of classification is required that maximizes the number of EAS candidates that can be correctly identified while keeping the false positive rate under 1\%.
One possible approach to this problem is to use a convolutional neural network (CNN) \cite{CNN}.

A CNN is a type of deep neural network (DNN) architecture that is popular in computer vision (CV) applications.
In these applications, the CNN derives a representation of the input image that is used in downstream regression and/or classification tasks \cite{lecun1998gradient}.
There are several qualities of CNN’s, and DNN’s in general, that make them attractive for the problem of onboard data classification.
First, common across all DNN’s is the training-inference dynamic.
During training, costly gradient calculations are required in order to optimize the many internal parameters of the model.
These gradient calculations are not required during inference, however, making predictions onboard our super pressure balloon relatively computationally lightweight \cite{copeland2016}.
Another advantage of DNN’s in our application is that they require minimal feature engineering and pre-processing of images.
The model will learn which features are most useful for differentiating between ``noise’’ and ``signal’’ events on its own with no prior knowledge \cite{lecun2015deep}, other than of course which events are ``noise’’ and which are ``signal’’.
Finally, CNN’s are advantageous for this application in particular because they are shift-invariant.
That is, the location of the signal in the camera should not impact the model’s ability to classify it correctly, all else being equal \cite{zhang1988shift, zhang1990parallel}.

In order to train the model, a large sample of simulated EAS was used.
Using the isotropic simulation methodology described in Section \ref{secs:EventRate}, 10,000 showers that passed the level 1 trigger condition described in Section \ref{secs:Trigger} were broken into mutually exclusive train (70\%), test (15\%) and evaluation (15\%) sets.
No special weighting was given to showers based on their energy or their appearance in our camera.
For the null data, or what the model was trained to distinguish EAS from, a combination of simulated noise and data from Mini-EUSO \cite{MiniEUSO} and EUSO-SPB1 \cite{SPB1} were used.
Concretely, our model is a 5-layer CNN model which derives a representation for an input image of size 24x24.
This representation is then used to make a binary prediction as to whether or not the image contains a signal.
The model was trained for 20 epochs using an Adam optimizer with a learning rate of 0.001, L2 penalty of 0.01, and dropout rate of 20\%.

\begin{figure}
\centering
\begin{minipage}{.47\textwidth}
  \centering
\includegraphics[width=\textwidth]{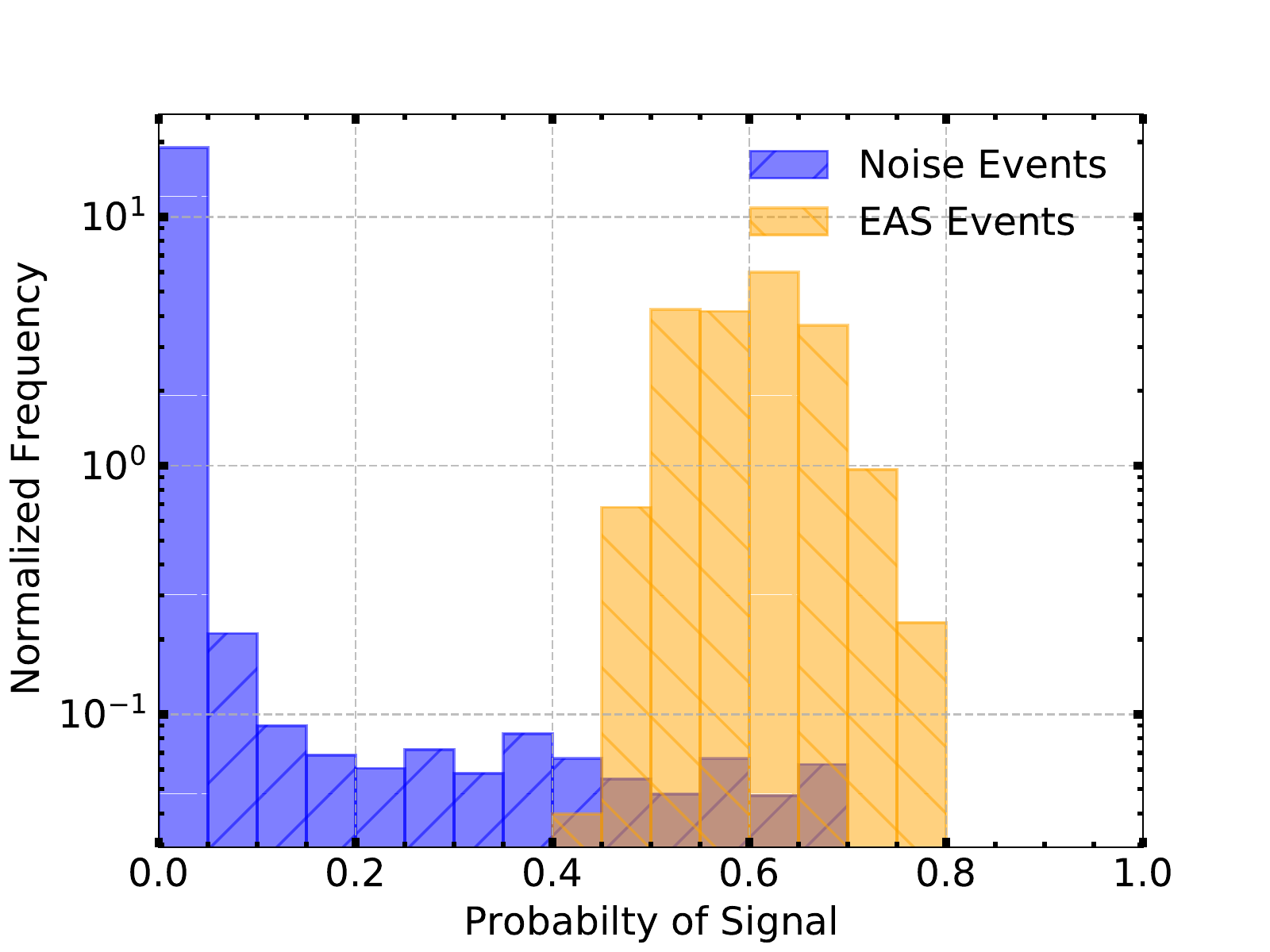}
\end{minipage}
\begin{minipage}{.47\textwidth}
  \centering
\includegraphics[width=\textwidth]{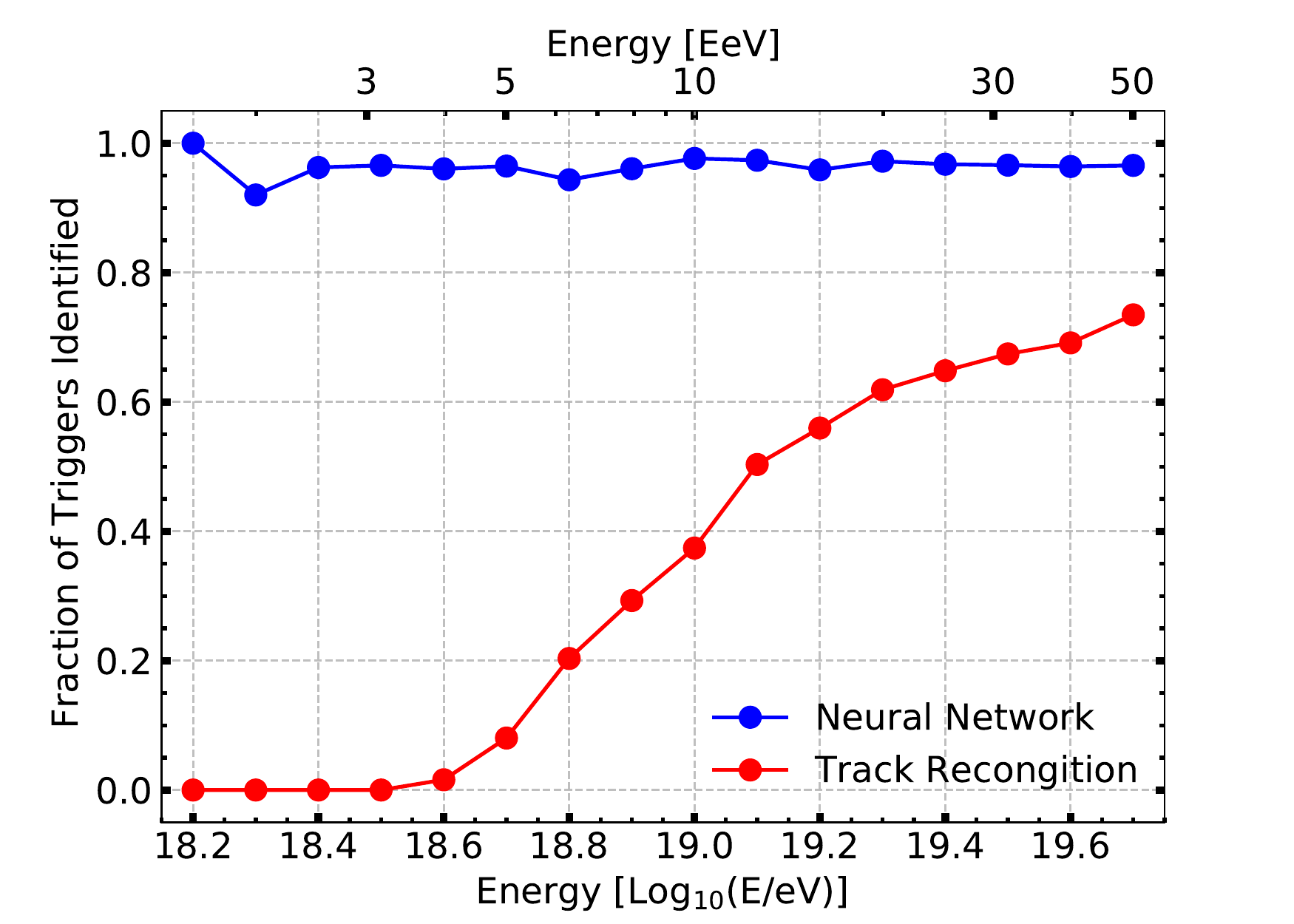}
\end{minipage}
\caption{Output of CNN, distribution of the probability of signal for noise and signal events (left).
Fraction of triggered events correctly identified as EAS by both the neural network approach and a traditional track recognition algorithm (right).
Events with a probability of signal >0.5 are considered correctly identified.}
\label{fig:CNNDist}
\end{figure}

The model reached a 97\% success rate overall correctly identifying the vast majority of EAS candidates across all energies.
Additionally, the model successfully identified noise events with an accuracy of 99.5\%.
The distribution of calculated probabilities of signal for EAS and noise is shown in Figure \ref{fig:CNNDist}.
In practice, the false positive rate (FPR) will most likely be higher as the noise triggers that EUSO-SPB2 records will be unique and the model cannot be trained using them prior to flight.
However, the relatively small size of a trained model leaves a possibility of it being updated mid-flight after a model is retrained using flight data if necessary.
Planned field tests prior to the flight will also allow for the model to be further refined.
An important feature of the CNN is that it is trained energy independently.
As a result, EAS at lower energies are recognized with similar accuracy to EAS at higher energies.
Given the nature of the UHECR energy spectrum, this leads to a much larger number of events that can be correctly identified compared with traditional track recognition algorithms, where the efficiency increases with energy.
This effect is illustrated in the left panel of Figure \ref{fig:CNNDist}.

\section{Conclusions}\label{secs:Conclusions}

By using extensive simulations carried out via the JEM-EUSO OffLine framework, the development of the EUSO-SPB2 FT onboard software has been informed and the expected performance has been characterized.
The detector is expected to measure $0.12\pm0.01\pm0.04$ events per hour of clear observation, with a peak energy sensitivity at $10^{18.6}$ eV.
A large fraction of these events, >95\% should be identifiable by onboard software and downloadable during the flight.
We expect a smaller fraction of events will be of sufficient quality to be reconstructed.
This should be verifiable during field tests prior to launch, after which the expectations may be changed appropriately.
The reconstruction accuracy will not provide competitive scientific measurements, in terms of directional resolution, when compared with ground based experiments that are subject to a different set of constraints.
Nonetheless, the goal of making the first observation and reconstruction of an UHECR-induced EAS from above via fluorescence is well within reach.

\textit{Acknowledgment:}
This work was partially supported by NASA Grant Number: 80NSSC18K0477.
This research used resources of the National Energy Research Scientific Computing Center (NERSC), a U.S. Department of Energy Office of Science User Facility operated under Contract No. DE-AC02-05CH11231.

\bibliographystyle{JHEP}
\bibliography{references}

\textwidth=16.0cm
\textheight=25cm

\noindent
{\Large\bf The JEM-EUSO Collaboration\\}
{\scriptsize (author-list as of July 1st, 2021)} \hspace{0.6cm}
{\scriptsize (version  \today{} \currenttime{})}
\vspace*{0.5cm}

\begin{sloppypar}
{\small \noindent
G.~Abdellaoui$^{ah}$,
S.~Abe$^{fq}$,
J.H.~Adams Jr.$^{pd}$,
D.~Allard$^{cb}$,
G.~Alonso$^{md}$,
L.~Anchordoqui$^{pe}$,
A.~Anzalone$^{eh,ed}$,
E.~Arnone$^{ek,el}$,
K.~Asano$^{fe}$,
R.~Attallah$^{ac}$
% ...
}
\end{sloppypar}
\vspace*{.3cm}

{ \footnotesize
\noindent
% ...
$^{ac}$ LPR at Dept. of Physics, Faculty of Sciences, University Badji Mokhtar, Annaba, Algeria \\
% ...
$^{ah}$ Telecom Lab., Faculty of Technology, University Abou Bekr Belkaid, Tlemcen, Algeria \\
% ...
$^{cb}$ APC, Univ Paris Diderot, CNRS/IN2P3, CEA/Irfu, Obs de Paris, Sorbonne Paris Cit\'e, France\\
% ...
$^{ed}$ Istituto Nazionale di Fisica Nucleare - Sezione di Catania, Italy\\
% ...
$^{eh}$ INAF - Istituto di Astrofisica Spaziale e Fisica Cosmica di Palermo, Italy\\
% ...
$^{ek}$ Istituto Nazionale di Fisica Nucleare - Sezione di Torino, Italy\\
$^{el}$ Dipartimento di Fisica, Universita' di Torino, Italy\\
% ...
$^{fe}$ Institute for Cosmic Ray Research, University of Tokyo, Kashiwa, Japan\\
% ...
$^{fq}$ Nihon University Chiyoda, Tokyo, Japan\\
% ...
$^{md}$ Universidad Polit\'ecnia de madrid (UPM), Madrid, Spain\\
% ...
$^{pd}$ University of Alabama in Huntsville, Huntsville, USA\\
$^{pe}$ Lehman College, City University of New York (CUNY), USA\\
% ...

}

\vspace*{0.5cm}

\vspace*{0.5cm}

{\small
\section*{Standard JEM-EUSO Acknowledgment in full-author papers}
This work was partially supported by Basic Science Interdisciplinary Research Projects of
RIKEN and JSPS KAKENHI Grant (22340063, 23340081, and 24244042), by
the Italian Ministry of Foreign Affairs	and International Cooperation,
by the Italian Space Agency through the ASI INFN agreements n. 2017-8-H.0 and n. 2021-8-HH.0,
by NASA award 11-APRA-0058 in the USA,
by the French space agency CNES,
by the Deutsches Zentrum f\"ur Luft- und Raumfahrt,
the Helmholtz Alliance for Astroparticle Physics funded by the Initiative and Networking Fund
of the Helmholtz Association (Germany),
by Slovak Academy of Sciences MVTS JEM-EUSO, by National Science Centre in Poland grants 2017/27/B/ST9/02162 and
2020/37/B/ST9/01821,
by Deutsche Forschungsgemeinschaft (DFG, German Research Foundation) under Germany's Excellence Strategy - EXC-2094-390783311,
by Mexican funding agencies PAPIIT-UNAM, CONACyT and the Mexican Space Agency (AEM),
as well as VEGA grant agency project 2/0132/17, and by by State Space Corporation ROSCOSMOS and the Interdisciplinary Scientific and Educational School of Moscow University "Fundamental and Applied Space Research".
}

\end{document}